# Verlässliche Software im 21. Jahrhundert

Stefan Wagner, Matthias Tichy, Michael Felderer, Stefan Leue
Universität Stuttgart, Universität Ulm, Universität Innsbruck, Universität Konstanz
stefan.wagner@informatik.uni-stuttgart.de, matthias.tichy@uni-ulm.de, michael.felderer@uibk.ac.at,

**Zusammenfassung**
Software ist Innovationstreiber in vielen verschiedenen Bereichen, von Cloud-Diensten über autonomes Fahren bis hin zu medizinischen Geräten und dem hochautomatisierten Aktienhandel. Allen diesen Bereichen ist gemeinsam, dass sie hohe Anforderungen an die Verlässlichkeit der Software stellen. Im Rahmen dieses Papiers diskutieren wir anhand von verschiedenen Beispielen neue Herausforderungen und mögliche Lösungswege, welche diese Bereiche an die Sicherstellung der Verlässlichkeit stellen. Wir unterscheiden bei diesen Herausforderungen sowohl Eigenschaften der Systeme, wie beispielsweise offene Systeme und ad-hoc Strukturen, als auch neue Aspekte der Verlässlichkeit, wie Nachvollziehbarkeit.

**Abstract**
Software is the main innovation driver in many different areas, like cloud services, autonomous driving, connected medical devices, and high-frequency trading. All these areas have in common that they require high dependability. In this paper, we discuss challenges and research directions imposed by these new areas on guaranteeing the dependability. On the one hand challenges include characteristics of the systems themselves, e.g., open systems and ad-hoc structures. On the other hand, we see new aspects of dependability like behavioral traceability.

## Innovationstreiber Software

Auch wenn es langsam klingt wie ein Klischee: Software durchdringt immer mehr Lebensbereiche. Kaum ein Lebensbereich ist mehr ohne Software vorstellbar. Selbst in Gebieten, die lange kaum von Softwareinnovationen profitiert haben, wie die Landwirtschaft, ist heute durch die fortschreitende Vernetzung und Verfügbarkeit von Rechenkapazität der Einsatz von Software normal geworden.

Diese umfassende Digitalisierung unserer Lebenswelt bietet nun die Basis für immer neue und immer komplexere Softwaresysteme, die verschiedene Bereiche verbinden, verknüpfen und dadurch völlig neue Möglichkeiten bieten. In dem Beispiel der Landwirtschaft können in Zukunft die Maschinen auf den Feldern mit Überwachungssystemen in den Ställen, Cloud-Diensten, wie detaillierten Wettervorhersagen, und den Systemen der Logistik und Produktion der Abnehmer eng zusammenarbeiten, um die Ernte genau zu optimieren. Gleichzeitig bringt die Digitalisierung und Vernetzung aber auch Risiken mit sich. Dienste können ausfallen, falsch arbeiten oder sogar angegriffen werden, was z.B. im schlimmsten Fall die Zerstörung der Ernte zur Folge hat. Mehr als jemals zuvor benötigen wir verlässliche Softwaresysteme, also Systeme, die zuverlässig und sicher sind. Worin liegen die Herausforderungen der Verlässlichkeit der Systeme der Zukunft?

Im Folgenden werden wir, nach einem kurzen Überblick zum Stand der Wissenschaft und Technik zu verlässlichen Software-Systemen, vier neue oder veränderte Verlässlichkeitsanforderungen anhand von jeweils einem repräsentativen Beispiel diskutieren:
- Nachvollziehbarkeit der Funktionsweise und Ergebnisse der Systeme
- Handhabung und Verteilung datenschutzrelevanter Daten und Verständnis der Wünsche der Nutzer bezüglich dieser Daten
- Zusammenspiel vieler Teilsysteme ad-hoc und mit minimaler Konfiguration
- Technische Sicherheit autonomer, adaptiver und lernender Teilsysteme

## Überblick zum Stand der Wissenschaft und Technik

Der Begriff der Verlässlichkeit (Dependability) (Avizienis et al., 2004) ist stark von der Domäne kritischer Systeme beeinflusst (siehe Abbildung 1). So sind die Attribute der Verlässlichkeit nach Avizienis et al. Verfügbarkeit

(Availability), Zuverlässigkeit (Reliability), Sicherheit (Safety), Vertraulichkeit (Confidentiality), Integrität (Integrity) und Wartbarkeit (Maintainability). Informationssicherheit (Security) betrachtet inbesondere Vertraulichkeit, Verfügbarkeit und Integrität. Der Begriff der Verlässlichkeit wird vor allem in den Domänen Luft- und Raumfahrt, Fahrzeugtechnik, Kernkraft, Schienenverkehr und Medizin genutzt.

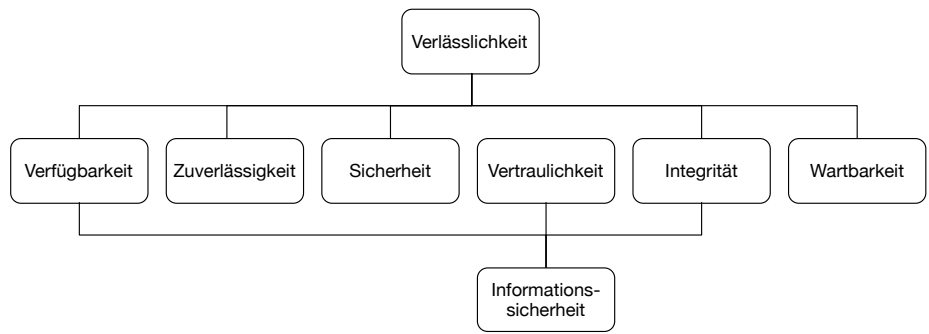

Abb. 1: Verlässlichkeit und Informationssicherheit nach (Avizienis et al., 2004)

Verlässlichkeit als ein Aspekt der Qualität von Systemen wird durch organisatorische, konstruktive und analytische Maßnahmen gesichert. Organisatorische Maßnahmen in den angesprochenen Domänen sind vor allem die extensive Anwendung von Standards: DO-178C in der Luftfahrt (RTCA, 2011), IEC 61508 allgemein zur funktionalen Sicherheit (IEC, 2010) und ihre Spezialisierungen EN 50128 im Schienenverkehr (EN, 2011) sowie ISO 26262 im Automobil (ISO, 2011). Diese Standards stellen umfangreiche Anforderungen an die Prozesse und Aktivitäten der Entwicklung und müssen durch Zertifizierungen nachgewiesen werden.

Konstruktive Maßnahmen sind beispielsweise die Nutzung von spezifischen Sprachen, zum Beispiel Rust (Matsakis und Klock, 2014), oder die nur eingeschränkte Nutzung von Features von C durch Anwendung der MISRA-Guidelines (MISRA, 2004). Aber auch die Anwendung bestimmter Vorgehensmodelle in der Softwareentwicklung zählen zu konstruktiven Maßnahmen. Für Systeme mit hohen Verlässlichkeitsanforderungen sind dabei nach wie vor lineare, schwergewichtige Modelle wie das V-Modell weit verbreitet. Mit der zunehmenden Verbreitung verlässlichkeitskritischer Systeme nimmt auch der Druck zu, die Iterationszyklen wie in der agilen Entwicklung zu verkürzen und die Qualitätssicherung weiter zu automatisieren.

Darüber hinaus existieren zahlreiche analytische Maßnahmen zur Gewährleistung der Verlässlichkeit. Beispielsweise muss eine Gefährdungsanalyse für das System während der kompletten Entwicklung sowie während des Betriebs durchgeführt werden (Storey, 1996). Als Technik werden, neben FMEA (Failure Mode and Effects Analysis) oder HAZOP (Hazard and Operability Study), beispielsweise Fehlerbaumanalysen genutzt, um zu identifizieren welche Gefahrensituation durch Fehler oder eine Kombination von Fehlern auftreten können. Die Nutzung mathematisch fundierter Methoden (formale Methoden) ist eine weitere analytische Maßnahme. Hier wird das Verhalten des Systems formal spezifiziert und dann hinsichtlich bestimmter Anforderungen analysiert. Formale Methoden werden etwa genutzt, um die Korrektheit eines Systems in Bezug auf Verlässlichkeitseigenschaften zu zeigen (zum Beispiel mit Model Checking auf Basis von Automaten (Bengtsson und Yi, 2004)) oder um die Verfügbarkeit des Systems mathematisch zu analysieren (zum Beispiel mit analytischen und simulationsbasierten Verfahren auf Markovketten) (Kwiatkowska et al., 2011). Formale Methoden spielen aber auch eine Rolle in Bezug auf klassische Qualitätssicherungstechniken wie dem Testen, wo formale Modelle als Grundlage für modellbasiertes Testen genutzt werden können, um Testfälle zu erzeugen (Utting and Legeard, 2010).

Diese verschiedenen Kategorien von Qualitätssicherungsmaßnahmen stehen natürlich nicht in Isolation, sondern hängen zusammen. So enthalten die oben angesprochenen Standards Anforderungen an den Einsatz von konstruktiven und analytischen Maßnahmen.

Grundlegende Annahme der meisten derzeit in der Praxis angewandten und in der Forschung existierenden Ansätze zur Analyse der Verlässlichkeit ist, dass die Systeme komplett bekannt sind und keine Verbindungen zur Außenwelt existieren (closed world) bzw. diese nicht relevant sind. Das bedeutet, dass diese auf neue Softwaresysteme, wie Software für High-Frequency-Trading oder vernetzte Cyber-physische Systeme, nicht zufriedenstellend angewendet werden können. Wenngleich die Software- und Systemarchitekturen dieser Systeme bereits recht gut beschrieben und analysiert werden können, stellen die Dynamik und die durch die verwendeten Regelungsalgorithmen erzeugten kontinuierlichen Zustandsräume besondere Herausforderungen für Analyse und Testverfahren dar. Weiterhin stellt die Größe und das diesen Systemen eigene nebenläufige Verhalten eine Herausforderung für die Anwendbarkeit und Skalierbarkeit aktueller Techniken dar. Lassen Sie uns dies an einigen konkreten Beispielen genauer betrachten.

## Nachvollziehbarkeit der Funktionsweise und Ergebnisse der Systeme – High-Frequency Trading

In der Geschäftswelt ist der Einsatz von hochkomplexer Software bereits häufig anzutreffen. An den Börsen betreibt Software das sogenannte High-Frequency Trading. Algorithmen reagieren innerhalb von Bruchteilen einer Sekunde auf Marktveränderungen und kaufen oder verkaufen entsprechende Wertpapiere. Das Geschehen ist für Menschen oft nicht mehr durchschaubar. So sind die Gründe und Ursachen des Flash Crashs von 2010 immer noch nicht eindeutig geklärt (Siedenbiedel, 2015). Ähnliches gilt für das Credit Scoring, in dem vollautomatisch die Kreditwürdigkeit einer Person bewertet wird. Beide Arten von Systemen haben großen gesellschaftlichen Einfluss und ihr Versagen hat starke negative Auswirkungen. Gleichzeitig sind die darin implementierten Entscheidungsprozesse so komplex und adaptiv, dass die Nachvollziehbarkeit für den Menschen verloren geht. Neben dem Wunsch, dass solche Systeme möglichst akkurate Entscheidungen treffen, kommt nun die Herausforderung hinzu, diese Entscheidungen auch verständlich und nachvollziehbar zu erklären, anstatt nur einfach den Credit Score darzustellen (Carrns, 2015).

Für das High-Frequency-Trading sind durchaus Aspekte der klassischen Verlässlichkeit relevant, zum Beispiel Verfügbarkeit und Zuverlässigkeit. So werden Anforderungen, dass Algorithmen nicht zu massiven Ausschlägen an Märkten führen und unterbrechungsfrei arbeiten, an solche Systeme gestellt. Hier könnten prinzipiell Techniken zur Analyse des Verhaltens dieser Systeme, etwa mit formalen Methoden, angewendet werden. Dadurch dass der Aktienmarkt offen ist, können beliebige und damit unbekannte und in ihrer Anzahl sich dynamisch ändernde Systeme interagieren. Selbst wenn die einzelnen Systeme korrekt Ihre Anforderungen erfüllen, könnte durch diese Interaktionen unerwünschtes und unerklärbares Verhalten des Gesamtsystems zustande kommen. Es müssen daher Techniken genutzt und entwickelt werden, die besser auf sich dynamisch ändernde Systeme angepasst sind.

Ein vielversprechender Ansatz sind Abstraktionstechniken basierend auf Modellen, welche helfen die Komplexität, Offenheit, Adaptivität und Nachvollziehbarkeit beherrschbar zu machen. Modellbasierte Ansätze sollen dabei sowohl generierende Modelle, beispielsweise für das modellbasierte Testen basierend auf Formalen Methoden, als auch prädiktive Modelle, basierend auf maschinellem Lernen, integrieren. Diese Abstraktionstechniken sollten es dann ermöglichen, etwa relevante Eigenschaften des Verhaltens von High-Frequency-Trading-Techniken abzubilden, um das Zusammenwirken von mehreren dieser Algorithmen auf einem offenen Markt mit zugehörigen Analysetechniken zu untersuchen. Eine Herausforderung ist hierbei, wie man zum einen von den konkreten Details der Algorithmen soweit abstrahieren kann, dass Geschäftsgeheimnisse gewahrt werden und die Analyse noch skaliert, aber zum anderen noch aussagekräftige Analyseergebnisse gewonnen werden können und ein Mensch die Ergebnisse noch nachvollziehen kann. Hier ergibt sich ein Lösungsraum, in dem verschiedene Algorithmen einen unterschiedlichen Trade-Off realisieren können.

## Handhabung datenschutzrelevanter Daten verständlich für die Nutzer – Health Tracker

So genannte Health Tracker werden immer beliebter und auch Teil von anderen Geräten, die wir häufig benutzen (Jüngling, 2015). So erfassen aktuelle Smartwatches typischerweise regelmäßig den Puls und die Anzahl der Schritte des Trägers. Dies vermittelt dem Nutzer ein Bild über seinen Gesundheitszustand und kann anspornen, mehr Bewegung in den Tagesablauf einzubauen. Inzwischen interessieren sich aber auch Krankenversicherungen für solche Health Tracker, da für sie der Einblick in den Gesundheitszustand eines Versicherungsnehmers höchst spannend ist: Sie würden gerne gesunde Lebensweisen mit Rabatten belohnen. Damit ergibt sich für die Endanwender die Abwägung, wie viel und welche Daten sie weitergeben wollen und in welcher Qualität. Dies ist sowohl für die Endanwender schwer zu verstehen und zu entscheiden, als auch technisch noch eine Herausforderung, die übertragenen Daten so zu verändern, dass die Datenschutz-Bedürfnisse der Endnutzer gewahrt bleiben.

Für das Beispiel der Health Tracker müssen analytische Ansätze entwickelt werden, die es nicht nur dem Anbieter von solchen Geräten, Ärzten, oder Experten bei Krankenkassen ermöglichen, die gesammelten Daten und deren Effekte, zum Beispiel auf Krankenkassenverträge, zu analysieren. Hier müssen Ansätze entstehen, die auch ein normaler Benutzer solcher Anwendungen versteht und die dem Nutzer für eine Vorhersehbarkeit und Nachvollziehbarkeit erklären, wie die Preisgabe seiner Daten Auswirkungen kurzfristig und langfristig auf Krankenkassenbeiträge, Zinsen bei einer Baufinanzierung oder Jobaussichten hat. Hier bieten sich auch What-If-Analysen an, bei denen der Benutzer durch systematische Variation von der Art und Genauigkeit der freigegebenen Daten erfährt, welche Auswirkungen diese Freigabe hat. Schlussendlich müssen dem Benutzer Möglichkeiten gegeben werden, die Einhaltung seiner Vorgaben zur Verwendung zu prüfen, bzw. diese Aufgabe einer Prüforganisation zu übertragen, ohne dass diese automatisch auf die Benutzerdaten selbst Zugriff erhält. Insgesamt soll dies dem Benutzer ermöglichen, informiert Entscheidungen hinsichtlich der Verwendung seiner persönlichen Daten zu treffen. Die Möglichkeit der Nachvollziehbarkeit und Erwartbarkeit sowie Prüfung durch den Benutzer ist auch eine zwingende Voraussetzung, um über ethisches Verhalten solcher Systeme diskutieren zu können.

## Zusammenspiel vieler Teilsysteme ad-hoc und mit minimaler Konfiguration – Smart Home

Ein weiteres Beispiel sind Systeme im Internet of Things. Dazu gehören mobile Systeme genauso wie fest installierte "Things" in der Heimautomatisierung. Gleichzeitig ist auch hier alles mit der Cloud vernetzt. Für Endanwender ist ein Konfigurationsaufwand nur sehr eingeschränkt zumutbar. Deshalb werden solche Internet-of-Things-Systeme ad-hoc entstehen und in großem Umfang zusammenarbeiten müssen, ohne vorher schon genau zu wissen, wer mit wem kooperieren muss. Die Orchestrierung dieser Systeme und Teilsysteme, so dass sie zuverlässig und performant für die Endnutzer ihren Dienst tun, stellt eine enorme Herausforderung dar.

In diesem Kontext ist auch die Forschung aus dem Bereich der selbst-adaptiven Systeme (de Lemos et al., 2017) relevant. Hier werden bereits Techniken entwickelt, um Systeme zu analysieren, die nicht komplett bekannt sind, und deren konkretes Zusammenarbeiten erst zur Laufzeit ad-hoc entsteht. Weiterhin müssen formale Analyseverfahren um Lernverfahren ergänzt werden, die zum Beispiel die tatsächlich erreichten Werte kontinuierlicher Systemparameter erlernen und somit das Testen bzw. die formale Verifikation von Regelungssystemen erleichtern. Allgemein muss also die Software-Verifikation (im weiteren Sinne) "vielseitiger" und adaptiver werden. Letztlich können die zur Entwicklungszeit verwendeten Software-Verifikationsmethoden durch Monitore ergänzt werden, die zur Laufzeit des Systems Abweichungen vom erwarteten Systemverhalten anzeigen.

## Technische Sicherheit autonomer, adaptiver und lernender Teilsysteme – Autonome Fahrzeuge und Roboter

Eng verwandt mit dem Internet of Things sind cyber-physische Systeme (CPS). Dort wird das Zusammenspiel von Software mit der physischen Welt durch elektrisch/elektronische und mechanische Komponenten betont. Beispielsweise sind Produktionsroboter traditionell in abgesperrten Bereichen tätig. Zur Verbesserung der Zusammenarbeit mit den Arbeitern in der Fabrik und damit zur Erhöhung der Produktivität werden diese Schranken aber nach und nach abgebaut. Noch drastischer sieht man diese Annäherung bei den Plänen für den Einsatz von Robotern in der Seniorenpflege in Japan (Dowideit, 2015). Über die Elektronik und Mechanik greifen die CPS direkt in die reale Welt ein und können somit Unfälle mit Schädigungen an Menschen auslösen. Gerade im Zusammenspiel mit den oben genannten Herausforderungen der Adaptivität und der ad-hoc zusammenarbeitenden Systeme wird es immer schwieriger, die Technische Sicherheit (Safety) zu gewährleisten.

In ähnlicher Weise wird Software in Zukunft die Kontrolle über das Autofahren übernehmen. Am Ende der Entwicklung zum autonomen Fahren wird das fahrerlose Fahrzeug stehen, in dem die Software über Routen, Fahrspurwechsel und die Reaktion auf Notfallsituationen verlässlich zu entscheiden hat. Es ist sogar möglich, dass Software über Leben und Tod entscheiden muss, etwa wenn die Wahl besteht, entweder die Passagiere im eigenen Fahrzeug oder Passanten möglicherweise lebensbedrohlich zu verletzen. Die Diskussion, was hier von einer verlässlich arbeitenden Software geleistet werden soll, wie also solche ethischen Dilemma im autonomen Fahren aufzulösen sind, hat bestenfalls gerade begonnen (Bonnefon et al., 2016).

Aktuelle Techniken sind hier nicht mehr direkt anwendbar. So ist im Kontext vernetzter Fahrzeuge das Gesamtsystem nicht bekannt und eine Analyse von nur Teilen des Gesamtsystems, die auf der Annahme beruhen, dass das Gesamtsystem bekannt ist, würde zu völlig verfälschten Resultaten führen. Die Verwendung maschineller Intelligenz ("deep learning") in autonomen Fahrzeugen zur Realisierung von essentiellen Fahrfunktionen führt ferner dazu, dass deren Verhalten nicht mehr einfach durch die Analyse von Entwurfsmodellen oder Programmcode erklärt werden kann. Es geht also das klassische ingenieurwissenschaftliche Paradigma des Entwicklungsprozesses als Abfolge Spezifikation-Implementierung-Verifikation verloren, da die Implementierung der Fahrfunktionen nicht mehr als Verfeinerung der Spezifikation verstanden werden kann und damit Code-basierte Analysemöglichkeiten nicht mehr zur Verfügung stehen. Da solche Systeme auch über keine detaillierten, sondern bestenfalls recht pauschale Spezifikationen verfügen (z.B. „transportiere die Passagiere unfallfrei und auf dem schnellsten Weg von A nach B") sind zudem klassische Verfahren zur Verifikation nicht weiter anwendbar. Es ergibt sich damit die Notwendigkeit, neuartige Verfahren zur Qualitätssicherung und ggf. zur Zertifikation zu entwickeln.

Ein vielversprechender Ansatz (STAMP – System-Theoretic Accident Model and Processes (Leveson, 2011)), der insbesondere aus der Funktionalen Sicherheit kommt, verwendet Systemtheorie. STAMP verbindet Unfälle und Gefährdungen nicht mehr mit der Wahrscheinlichkeit eines Fehlers in einer Komponente des Systems, sondern setzt den Fokus auf das Zusammenspiel der Komponenten im ganzen soziotechnischen System, welches insbesondere auch die beteiligten Menschen beinhaltet. Das Ergebnis einer STAMP-Analyse sind „unsichere Szenarien", in denen die Regelung des Systems fehlschlägt. Eine Beteiligung der Menschen ist beispielsweise durch den bidirektionalen Transfer der Kontrolle zwischen autonomen Systemen und ihren menschlichen Nutzern erforderlich (Wahlster, 2017). Ein weiteres Beispiel sind hybride Mensch-Technik-Systeme, in denen hybride Teams aus mehreren autonomen Systemen und mehreren Menschen lösen eine Aufgabe gemäß ihren spezifischen Fähigkeiten gemeinsam als kollektive Intelligenz.

Eine weitere wichtige Verbesserung in der Softwareentwicklung in den letzten Jahren, um Komplexität und Adaptivität von Systemen beherrschbar zu machen, ist sicherlich die Automatisierung von Tests und deren Einbindung in eine kontinuierliche Softwareentwicklung. Diese kontinuierliche Qualitätskontrolle erlaubt eine sehr schnelle Rückmeldung zu potentiellen Problemen in der Software. Speziell für verlässliche Softwaresysteme wird es wichtig sein, den Einsatz der kontinuierlichen Qualitätskontrolle noch weiter auszubauen und mit formalen Verifikationstechniken wie Model Checking zu kombinieren, um komplexe und adaptive Systeme wie moderne cyber-physische Systeme schnell validieren und verifizieren zu können und dadurch Feedback-Zyklen kurz zu halten.

## Zusammenfassung und Ausblick

Wir brauchen weiterhin die bisher verwendeten Qualitätssicherungsmaßnahmen zur Sicherstellung von Verlässlichkeit, wie Standards, Fehleranalysetechniken oder formale Methoden. Gleichzeitig zeigt die Praxis, dass existierende Techniken die neuen Herausforderungen an Verlässlichkeit nur unzureichend abdecken. Insbesondere fehlen Ansätze, welche neue Herausforderungen an die Komplexität, Offenheit, Adaptivität und Nachvollziehbarkeit solcher Systeme hinreichend und integriert adressieren.

Dieser Artikel kann nicht alle interessanten Lösungsansätze abdecken. Aber wir sind davon überzeugt, dass durch die Kombination und Weiterentwicklung der genannten Ansätze, wie beispielsweise kontinuierlicher Entwicklung mit kurzen Feedback-Zyklen, basierend auf systemtheoretischer Analyse und ergänzt mit gezielten formalen Beschreibungen und Verifikation, verlässliche Software für das 21. Jahrhundert erreicht werden kann.

Insgesamt muss sich die Informatik stärker der Tatsache stellen, dass sie in allen Gebieten des menschlichen Lebens angekommen ist. Dies bedeutet, dass eine noch weitaus stärkere Verknüpfung der Informatik mit anderen Disziplinen notwendig ist. Offensichtlich sind Bezüge zu anderen Ingenieurdisziplinen wie Maschinenbau oder Elektrotechnik. Dort ist bereits viel passiert, aber auch immer noch viel Aufholbedarf, wenn man sich die Unterschiede in der heutigen Entwicklung von Mechanik, Elektronik und Software ansieht. Dies beinhaltet nicht nur die Entwicklung gemeinsamer Techniken, sondern auch deren Verbesserung der Anwendbarkeit in der Praxis für die Ingenieure der verschiedenen Disziplinen. Aber insbesondere auch Medizin, BWL, VWL, Psychologie, Soziologie, Jura und Ethik werden enge Partner der Informatik werden müssen, um die menschlichen und wirtschaftlichen Aspekte besser zu verstehen und in Lösungen integrieren zu können.

## Literatur